\documentclass[10pt,final,journal,a4paper,oneside,twocolumn]{IEEEtran}

\usepackage{cite}
\usepackage{ulem}
\usepackage{soul,color}
\usepackage{srcltx}
\usepackage{multirow}
\usepackage[tbtags,sumlimits,nointlimits,reqno]{amsmath}
\allowdisplaybreaks[4]
\usepackage{graphicx}
\usepackage{stfloats}
\usepackage{subfigure}
\usepackage{psfrag}
\usepackage{color}
\usepackage{amssymb}
\usepackage{epsfig}
\usepackage{epstopdf}

\begin{document}

\title{Capacitor-Loaded Spoof Surface Plasmon (SSP)\\ for Flexible Dispersion Control and \\ High-Selectivity Filtering }
\author{Xiao-Lan~Tang,
        Qingfeng~Zhang,~\IEEEmembership{Senior~Member,~IEEE,}
        Sanming~Hu,~\IEEEmembership{Senior~Member,~IEEE,}
        Abhishek~Kandwal,
        Tongfeng~Guo,~\IEEEmembership{Student Member,~IEEE,}
        and~Yifan~Chen,~\IEEEmembership{Senior Member,~IEEE,}

\thanks{This work has been submitted to the IEEE for possible publication. Copyright may be transferred without notice, after which this version may no longer be accessible.}
\thanks{
X.-L. Tang, Q. Zhang, A. Kandwal, T. Guo and Y. Chen are with the Department of Electrical and Electronic Engineering, South University of Science and Technology of China, Guangdong 518055, China. X.-L. Tang is also with the State Key Laboratory of Millimeter Waves, Southeast University, Nanjing 210096, China. (e-mail: zhang.qf@sustc.edu.cn).
}
\thanks{S. Hu is with the State Key Laboratory of Millimeter Waves, Southeast University, Nanjing 210096, China.}
}
\maketitle

\begin{abstract}

This letter proposes a new spoof surface plasmon transmission line (SSP-TL) using capacitor loading techniques. This new SSP-TL features flexible and reconfigurable dispersion control and highly selective filtering performance without resorting to configuration change. Moreover, it requires much smaller line width than the conventional SSP-TLs for achieving a extremely slow wave (or a highly confined field), which is quite useful for a compact system. To illustrate the design principle, several examples are designed within the frequency range of $2-8$~GHz. Both numerical and experimental results are given in comparison with the conventional SSP-TL. It is demonstrated that the proposed technique provides a better performance in size reduction and dispersion reconfigurability.

\end{abstract}


\IEEEpeerreviewmaketitle

\section{Introduction}

Surface plasmons, originally proposed and applied in optics, are propagating waves along the interface between a metal and a dielectric medium~\cite{ref1}. These waves, decaying exponentially in the direction perpendicular to the interface, exhibit a highly confined and localized field near the surface. In 2004, Pendry et al. has demonstrated that a similar surface plasmonic phenomenon, known as spoof surface plasmon (SSP), can be observed at microwave frequencies by perforating the conductive surface using parametrical holes~\cite{ref2}. Since then, SSPs are widely explored in low-crosstalk transmission lines~\cite{ref3,ref4,ref5,ref7}, filters~\cite{ref6}, leaky-wave antennas~\cite{ref8,ref9} and so on, due to their potential applications in millimeter-wave and terahertz circuits.

It has been demonstrated in the recent work~\cite{ref3} that the dispersion properties of SSP transmission line (SSP-TL) can be adjusted by modifying the geometrical dimensions and/or dielectric material parameters. The working principle lies in that the surface corrugation and dielectric parameters determine the effective capacitance along the surface and hence the electrical field confinement and dispersion property. However, a highly confined field usually requires a high-depth corrugation and hence a large line width. Such SSP-TLs, typically used for high-density transmission lines with low crosstalk, would lose its benefit on compactness. To overcome this drawback, we propose a capacitor-loaded SSP-TL. It features a small line width but with a strongly confined field, which significantly reduces the size of the whole system. Moreover, this new SSP-TL exhibits the benefits on flexible control of the dispersion curve without resorting to changing the geometrical configurations. It can be potentially extended to reconfigurable SSP-TLs by replacing the static capacitors with tunable ones. Typical applications may include highly selective filters with reconfigurable cutoff frequencies.

\section{Principle}

 \begin{figure}[!t]
   \centering
   \psfrag{a}[c][c]{\footnotesize $a$}
   \psfrag{w}[c][c]{\footnotesize $w$}
   \psfrag{d}[c][c]{\footnotesize $d$}
   \psfrag{h}[c][c]{\footnotesize $h$}
   \psfrag{1}[c][c]{\footnotesize $C_1$}
   \psfrag{0}[c][c]{\footnotesize $\Delta C$}
   \psfrag{2}[c][c]{\footnotesize $C_2$}
   \psfrag{3}[c][c]{\footnotesize $C_3$}
   \psfrag{4}[c][c]{\footnotesize $L_2$}
   \psfrag{5}[c][c]{\footnotesize $L_1$}
   \psfrag{6}[c][c]{\footnotesize $L_3$}
   \psfrag{7}[c][c]{\footnotesize In}
   \psfrag{8}[c][c]{\footnotesize Out}
   \psfrag{x}[c][c]{\footnotesize (a)}
   \psfrag{y}[c][c]{\footnotesize (b)}
   \psfrag{p}[c][c]{\footnotesize $z$}
   \psfrag{q}[c][c]{\footnotesize $y$}
   \psfrag{s}[c][c]{\footnotesize $x$}
   \includegraphics[width=6.5cm]{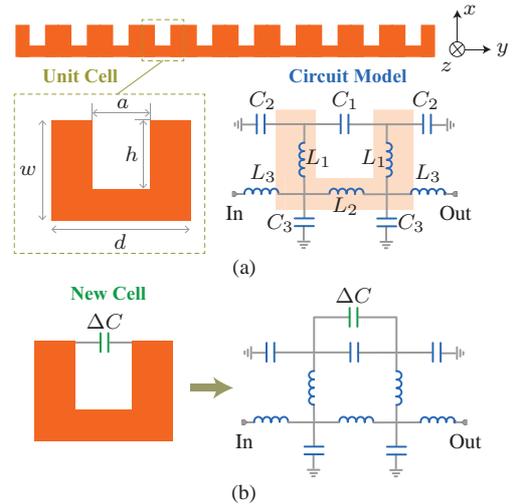}
   \caption{SSP-TL using surface corrugation: (a) conventional one, (b) proposed capacitor-loaded one.}\label{fig:fig1}
\end{figure}

Conventional SSP-TL is realized by corrugating a metallic strip with subwavelength periodic slots, as shown in Fig.~\ref{fig:fig1}(a), where the parameters $d$, $h$, $a$ and $w$ denote the periodic pitch, the slot depth, the slot width and the line width, respectively. The corrugation provides a capacitive loading along the strip, and hence supports a bounded mode confined to the surface, allowing to manipulate electromagnetic waves at subwavelength regime. The unit cell of SSP-TL is modeled by an equivalent circuit, as shown on the right side of Fig.~\ref{fig:fig1}(a)~\cite{ref3}, where $L_1$, $L_2$ and $L_3$ are attributed to the magnetic field excited by the currents along the strip units, $C_2$ and $C_3$ models the electric field pointing to the infinity (in $x$ direction), and $C_1$ represents the local electric field excited between the gap (mostly in $y$ direction).

\begin{figure}[!t]
   \centering
   \psfrag{a}[c][c]{\footnotesize $\beta d/\pi$}
   \psfrag{b}[c][c]{\footnotesize Frequency (GHz)}
   \psfrag{c}[r][c]{\footnotesize Air Line}
   \psfrag{d}[l][c]{\footnotesize CST: $h=3$~mm}
   \psfrag{e}[l][c]{\footnotesize CST: $h=5$~mm}
   \psfrag{f}[l][c]{\footnotesize CST: $h=8$~mm}
   \psfrag{g}[l][c]{\footnotesize CST: $h=11$~mm}
   \psfrag{p}[l][c]{\footnotesize E.C: $\Delta C=0$ pF}
   \psfrag{q}[l][c]{\footnotesize E.C: $\Delta C=0.1$ pF}
   \psfrag{r}[l][c]{\footnotesize E.C: $\Delta C=0.2$ pF}
   \psfrag{s}[l][c]{\footnotesize E.C: $\Delta C=0.5$ pF}
   \includegraphics[width=7.7cm]{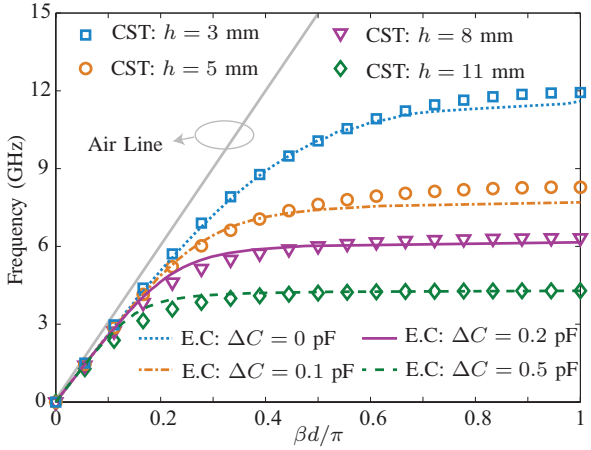}
   \caption{Dispersion Curves of (a) conventional SSP-TL and (b) capacitor-loaded SSP-TL. (E.C: Equivalent Circuit)}\label{fig:dispersion}
\end{figure}

To analyze the unit cell of Fig.~\ref{fig:fig1}(a), one adds a periodic boundary condition on both sides of the unit cell, and subsequently computes the dispersion curve using the software CST Microwave Studio. One may consider an example, in which the substrate is Rogers 4003C ($\epsilon_r=3.38$, $\tan\delta=0.0027$) with thickness of $1.52$~mm, and the strip dimensions in Fig.~\ref{fig:fig1}(a) are $a=1$~mm, $w=5$~mm, $d=5$~mm and $h=3$~mm. The calculated dispersion curve is shown as the blue squares in Fig.~\ref{fig:dispersion}. Note that it exhibits a typical SSP response, which stays close to the air line at low frequencies but goes away at high frequencies. The circuit parameters of Fig.~\ref{fig:fig1}(a) are calculated as $L_1=0.8$~nH, $L_2=0.14$~nH, $L_3=0.2$~nH, $C_1=22.5$~fF, $C_2=161$~fF and $C_3=178$~fF, by using the approach provided in~\cite{ref3}. One may further calculate the dispersion curve of the equivalent circuit to verify its validity. One firstly computes the scattering parameters ($S_{11}$ and $S_{21}$) of the circuit model, and subsequently obtains the dispersion relation by
\begin{equation}\
         \beta=\Im\left[ \frac{1}{d}\cosh^{-1}\left(\dfrac{1-S_{11}^2+S_{21}^2}{2S_{21}}\right)\right],~\label{eq:beta}
        \end{equation}
\noindent where $\Im[.]$ denotes the operation of taking the imaginary part. It should be noted that the above equation is obtained under the condition of symmetry ($S_{11}=S_{22}$) and reciprocity ($S_{21}=S_{12}$), and hence valid for symmetrical and reciprocal cases only. The computed dispersion curve using~\eqref{eq:beta} is displayed as the blue dotted line in Fig.~\ref{fig:dispersion}, which well follows the one obtained by CST full-wave simulation.

To increase the field confinement or localization, one should further bend the dispersion curve away from the air line. One way to achieve this is to increase the corrugation depth $h$, which is well validated by the dispersion curves of Fig.~\ref{fig:dispersion}. When $h$ increases from $3$~mm to $11$~mm, the dispersion curves gradually bend away from the air line, giving rise to a much slower trapped wave. Note that, in the case of a high corrugation depth (e.g. $h=11$~mm), the dispersion curve becomes almost flat when it is close to the cutoff frequency. This frequency, in analogy to the plasma frequency of surface plasmons in optics, is inversely proportional to the corrugation depth $h$. Also, the group velocity around this frequency is extremely small, which gives rise to a strongly localized wave in analogy to the strong light-electron interaction in optics. The small group velocity also leads to a sharp magnitude response, which is very useful in high-selectivity filters.

Although a high corrugation depth $h$ is good for field confinement, it inevitably increases the line width $w$ that is not preferred in a compact system. To overcome this drawback, we propose the capacitor-loaded SSP-TL in Fig.~\ref{fig:fig1}(b). The additional loaded capacitors along the surface brings in an extra capacitance, in equivalence to the extra capacitance brought by the increased corrugation depth in conventional SSP-TLs. Therefore, the proposed SSP-TL increases the field confinement capability and meanwhile maintains a small line width. The capacitor-loaded unit cell can be modeled by adding the extra capacitance $\Delta C$ to the original circuit, as shown in Fig.~\ref{fig:fig1}(b). One may use this circuit model to obtain the dispersion curves in Fig.~\ref{fig:dispersion}. Note that, the SSP-TL loaded with different capacitances achieves almost the same dispersion curves as the conventional SSP-TL with different depths.

\section{Experimental Validation}

\begin{figure}[!t]
  \centering
    \psfrag{a}[c][c]{\footnotesize (a)}
  \psfrag{b}[c][c]{\footnotesize (b)}
  \psfrag{x}[c][c]{\footnotesize \textcolor[rgb]{1.00,1.00,1.00}{$x$}}
  \psfrag{y}[c][c]{\footnotesize \textcolor[rgb]{1.00,1.00,1.00}{$y$}}
  \psfrag{z}[c][c]{\footnotesize \textcolor[rgb]{1.00,1.00,1.00}{$z$}}
  \psfrag{g}[c][c]{\footnotesize Capacitor-Loaded SSP}
  \psfrag{h}[c][c]{\footnotesize Conventional SSP}
    \psfrag{c}[l][c]{\footnotesize \textcolor[rgb]{1.00,1.00,0.00}{Chip Capacitor}}
  \psfrag{d}[c][c]{\footnotesize \textcolor[rgb]{0.50,1.00,0.00}{CPW-to-SSP}}
  \psfrag{e}[c][c]{\footnotesize \textcolor[rgb]{0.50,1.00,0.00}{Transition}}
  \includegraphics[width=8cm]{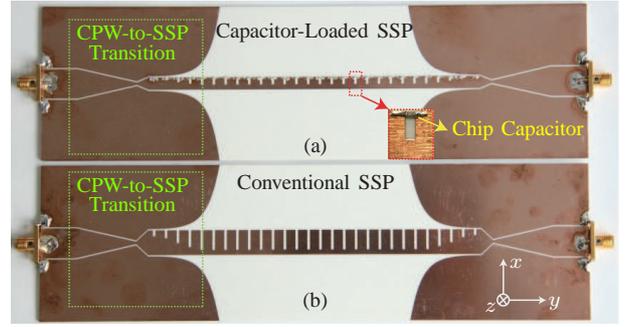}
  \caption{Fabricated prototypes of the capacitor-loaded SSP-TL and conventional SSP-TL.}\label{fig:fig2}
\end{figure}

To validate the proposed capacitor-loaded SSP-TL, we experimentally implement the one with $\Delta$C of 0.2~pF on a 1.52-mm-thick Rogers 4003C substrate, in correspondence with the purple-line dispersion curve of Fig.~\ref{fig:dispersion}. In comparison, we also implement the conventional SSP-TL with a corrugation depth $h=8$~mm, exhibiting a similar dispersion curve to the proposed SSP-TL. Fig.~\ref{fig:fig2} shows the fabricated prototypes of the two SSP-TLs. Tapered lines are employed for a smooth transition and mode conversion between the $50\Omega$-CPW line (width: $12$~mm, gap: $0.5$~mm) and SSP~\cite{ref4}. As shown in the zoomed figure of Fig.~\ref{fig:fig2}(a), the proposed SSP-TL is loaded with $0.2$~pF chip capacitors (Murata GJM1555C1HR20WB01D). The line parameters are $w=5$~mm, $h=3$~mm, $a=1$~mm, and $d=5$~mm. In contrast, the conventional SSP-TL has a much wider line width (i.e. $w=10$~mm), due to the increased corrugation depth $h=8$~mm. Thus, the capacitor-loaded SSP-TL reduces the line width by half.

\begin{figure}[!t]
  \begin{center}
  \psfrag{a}[c][c]{\footnotesize Frequency (GHz)}
  \psfrag{h}[c][c]{\footnotesize \textcolor[rgb]{0.50,0.50,0.50}{\textbf{Wide Stop Band}}}
  \psfrag{b}[c][c]{\footnotesize $|S_{21}|$ (dB)}
  \psfrag{c}[c][c]{\footnotesize $|S_{11}|$ (dB)}
  \psfrag{d}[l][c]{\footnotesize $|S_{21}|$ of Fig.~\ref{fig:fig2}(a)}
  \psfrag{e}[l][c]{\footnotesize $|S_{21}|$ of Fig.~\ref{fig:fig2}(b)}
  \psfrag{f}[l][c]{\footnotesize $|S_{11}|$ of Fig.~\ref{fig:fig2}(a)}
  \psfrag{g}[l][c]{\footnotesize $|S_{11}|$ of Fig.~\ref{fig:fig2}(b)}
  \includegraphics[width=8cm]{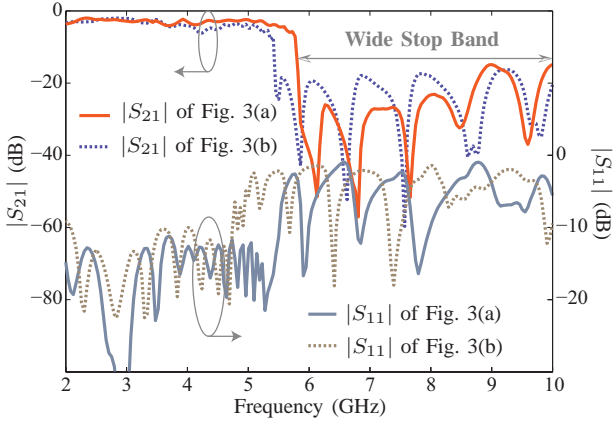}
  \end{center}
  \caption{The measured scattering parameters of the two fabricated SSP-TL prototypes in Fig.~\ref{fig:fig2}.}\label{fig:fig4}
\end{figure}

The two fabricated prototypes are measured using an Agilent PNA network analyzer (E5071C) within the frequency range $2-10$~GHz. The measured responses of the scattering parameters are shown in Fig.~\ref{fig:fig4}. Note that the capacitor-loaded SSP-TL has a similar response as the conventional SSP-TL, in spite of a slight discrepancy in the cutoff frequency, possibly due to the tolerances brought by PCB fabrication and the commercial chip capacitors ($\pm 0.05$~pF). Also note that, the proposed SSP-TL has a very sharp magnitude response around $5.8$~GHz, which exhibits a highly selective filtering feature. It also exhibits a wide stop band from $5.8$~GHz to $10$~GHz.

\begin{figure}[!t]
  \centering
  \psfrag{a}[c][c]{\footnotesize (a)}
  \psfrag{b}[c][c]{\footnotesize (b)}
  \psfrag{e}[l][c][0.7]{$\Delta z=10.9$ mm}
  \psfrag{u}[c][c]{\footnotesize $\Delta z$}
  \psfrag{f}[l][c][0.7]{$x=-6$ mm}
  \psfrag{h}[l][c][0.7]{$\Delta z=16.7$ mm}
  \psfrag{g}[l][c][0.7]{$x=-11$ mm}
  \psfrag{c}[c][c]{\footnotesize \textcolor[rgb]{1.00,1.00,1.00}{Proposed [Fig.~\ref{fig:fig2}(a)]}}
  \psfrag{d}[c][c]{\footnotesize \textcolor[rgb]{1.00,1.00,1.00}{Conventional [Fig.~\ref{fig:fig2}(b)]}}
  \includegraphics[width=8.6cm]{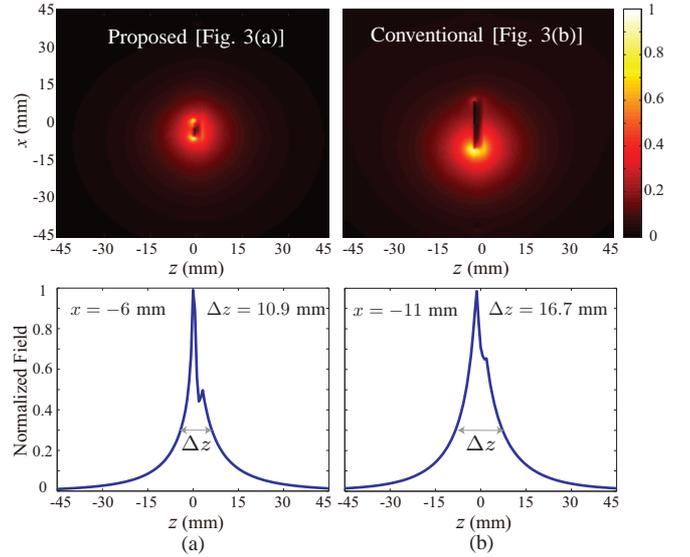}
  \caption{Cross-section distribution of the electric field of the SSP-TLs in (a) Fig.~\ref{fig:fig2}(a) and (b) Fig.~\ref{fig:fig2}(b) at $4$~GHz.}\label{fig:fig3}
\end{figure}

To compare the field confinement capability of the two SSP-TLs, their normalized electric fields in the cross section ($x-z$ plane) at $4$~GHz are plotted in Fig.~\ref{fig:fig3}. Note from the 2D distributions that the proposed SSP-TL has a better field confinement. To quantify this improvement, we plot the 1D $z$-distribution across the maximum point. Then one may define the confined region using the width when the field drops to $0.3$ (about $-10$~dB) of the maximum strength. Note that this width is $10.9$~mm in the proposed SSP-TL, and is $16.7$~mm in the conventional one. Thus, the confinement capability, inversely proportional to the confined region, is improved by $53\%$.

\begin{figure}[!t]
  \centering
  \psfrag{a}[c][c]{\footnotesize Frequency (GHz)}
  \psfrag{h}[c][c]{\footnotesize \textcolor[rgb]{0.50,0.50,0.50}{\textbf{Reconfigurable}}}
  \psfrag{g}[c][c]{\footnotesize \textcolor[rgb]{0.50,0.50,0.50}{\textbf{Operation Band}}}
  \psfrag{b}[c][c]{\footnotesize $|S_{21}|$ (dB)}
  \psfrag{c}[l][c]{\footnotesize $\Delta C=0.1$~pF}
  \psfrag{d}[l][c]{\footnotesize $\Delta C=0.2$~pF}
  \psfrag{e}[l][c]{\footnotesize $\Delta C=0.3$~pF}
  \psfrag{f}[l][c]{\footnotesize $\Delta C=0.5$~pF}
 \includegraphics[width=7.5cm]{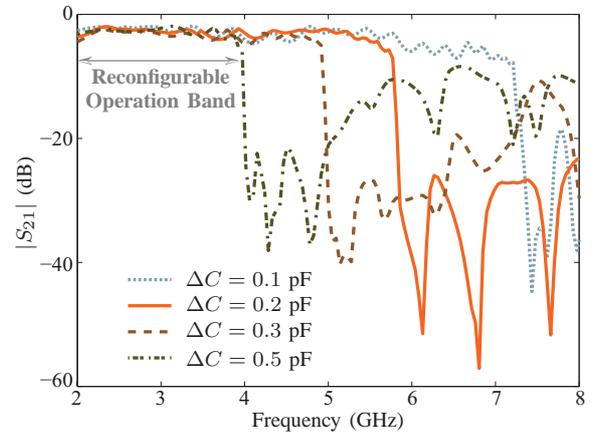}
  \caption{Measured magnitude responses of the fabricated prototype of Fig.~\ref{fig:fig2}(a) with different loading capacitances.}\label{fig:fig5}
\end{figure}

The proposed SSP-TL loaded with different capacitances are also fabricated and measured. The scattering parameters are shown in Fig.~\ref{fig:fig5}. Note that the cutoff frequency (in analogy to the plasma frequency in optics) and the operation band are completely reconfigurable by changing the capacitances. High-selectivity feature is still preserved while reconfiguring the operation band. The static capacitors may be replaced by tunable capacitors to make it truly reconfigurable, which will be further explored in our future works.

\section{Conclusion}

In this paper, a capacitor-loaded SSP-TL has been proposed. It featured reconfigurability in the dispersion control. It also exhibits a small line width in comparison with the conventional SSP-TL. The principle was illustrated and several experimental examples were provided. The results completely validated the proposed SSP-TLs.

\bibliographystyle{IEEEtran}
\bibliography{Mybib}

\end{document}